\documentclass[sigconf]{acmart}
\usepackage{tabularx, booktabs} 
\usepackage{graphicx, subcaption}

\AtBeginDocument{%
  \providecommand\BibTeX{{%
    \normalfont B\kern-0.5em{\scshape i\kern-0.25em b}\kern-0.8em\TeX}}}

\setcopyright{acmcopyright}
\copyrightyear{2018}
\acmYear{2018}
\acmDOI{10.1145/1122445.1122456}

\acmConference[Woodstock '18]{Woodstock '18: ACM Symposium on Neural
  Gaze Detection}{June 03--05, 2018}{Woodstock, NY}
\acmBooktitle{Woodstock '18: ACM Symposium on Neural Gaze Detection,
  June 03--05, 2018, Woodstock, NY}
\acmPrice{15.00}
\acmISBN{978-1-4503-XXXX-X/18/06}

\begin{document}
\title{A framework for constructing a huge name disambiguation dataset: algorithms, visualization and human collaboration}
\newcommand{\datasetName}{\textbf{WhoisWho}}
\newcommand{\placeholder}{???}
\newcommand{\vpara}[1]{\vspace{0.05in}\noindent\textbf{#1 }}

\author{Zhuoyue Xiao}
\email{joeyxiao317@gmail.com}
\affiliation{%
  \institution{Tsinghua University}
  \streetaddress{30 Shuangqing Rd}
  \city{Haidian Qu}
  \state{Beijing Shi}
  \country{China}}

\author{Yutao Zhang}
\authornote{Both authors contributed equally to this research.}
\email{yt-zhang13@mails.tsinghua.edu.cn}
\affiliation{%
  \institution{Tsinghua University}
  \streetaddress{30 Shuangqing Rd}
  \city{Haidian Qu}
  \state{Beijing Shi}
  \country{China}}

\author{Bo Chen}
\email{bochen@ruc.edu.cn}
\authornotemark[1]
\affiliation{%
  \institution{Renmin University of China}
  \city{Haidian Qu}
  \state{Beijing Shi}
  \country{China}}
  
\author{Xiaozhao Liu}
\email{xiaozhao.liu@duke.edu}
\affiliation{%
  \institution{Duke University}}
  
\author{Jie Tang}
\email{jietang@tsinghua.edu.cn}
\affiliation{%
  \institution{Tsinghua University}
  \streetaddress{30 Shuangqing Rd}
  \city{Haidian Qu}
  \state{Beijing Shi}
  \country{China}}

\begin{abstract}
We present a manually-labeled Author Name Disambiguation(AND) Dataset called \datasetName, which consists of 399,255 documents and 45,187 distinct authors with 421 ambiguous author names. To label such a great amount of AND data of high accuracy, we propose a novel annotation framework where the human and computer collaborate efficiently and precisely. Within the framework, we also propose an inductive disambiguation model to classify whether two documents belong to the same author. We evaluate the proposed method and other state-of-the-art disambiguation methods on \datasetName. The experiment results show that: (1) Our model outperforms other disambiguation algorithms on this challenging benchmark. (2) The AND problem still remains largely unsolved and requires more in-depth research. We believe that such a large-scale benchmark would bring great value for the author name disambiguation task. We also conduct several experiments which proves our annotation framework could assist annotators to make accurate results efficiently and eliminate wrong label problems made by human annotators effectively.
\end{abstract}



\begin{teaserfigure}
  \includegraphics[width=\textwidth]{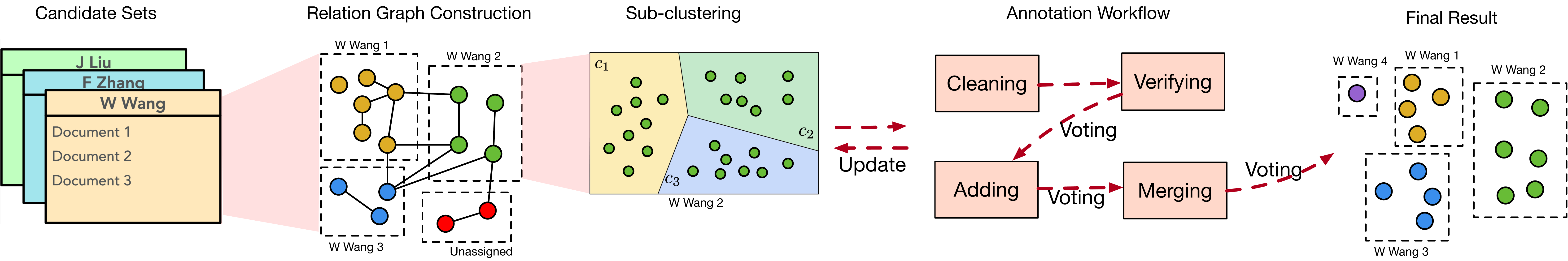}
  \caption{An overview of proposed annotation framework.}
  \label{fig:overview}
\end{teaserfigure}

\maketitle
\section{Introduction}
\begin{figure}[t]
	\centering
	\includegraphics[width=7.0cm]{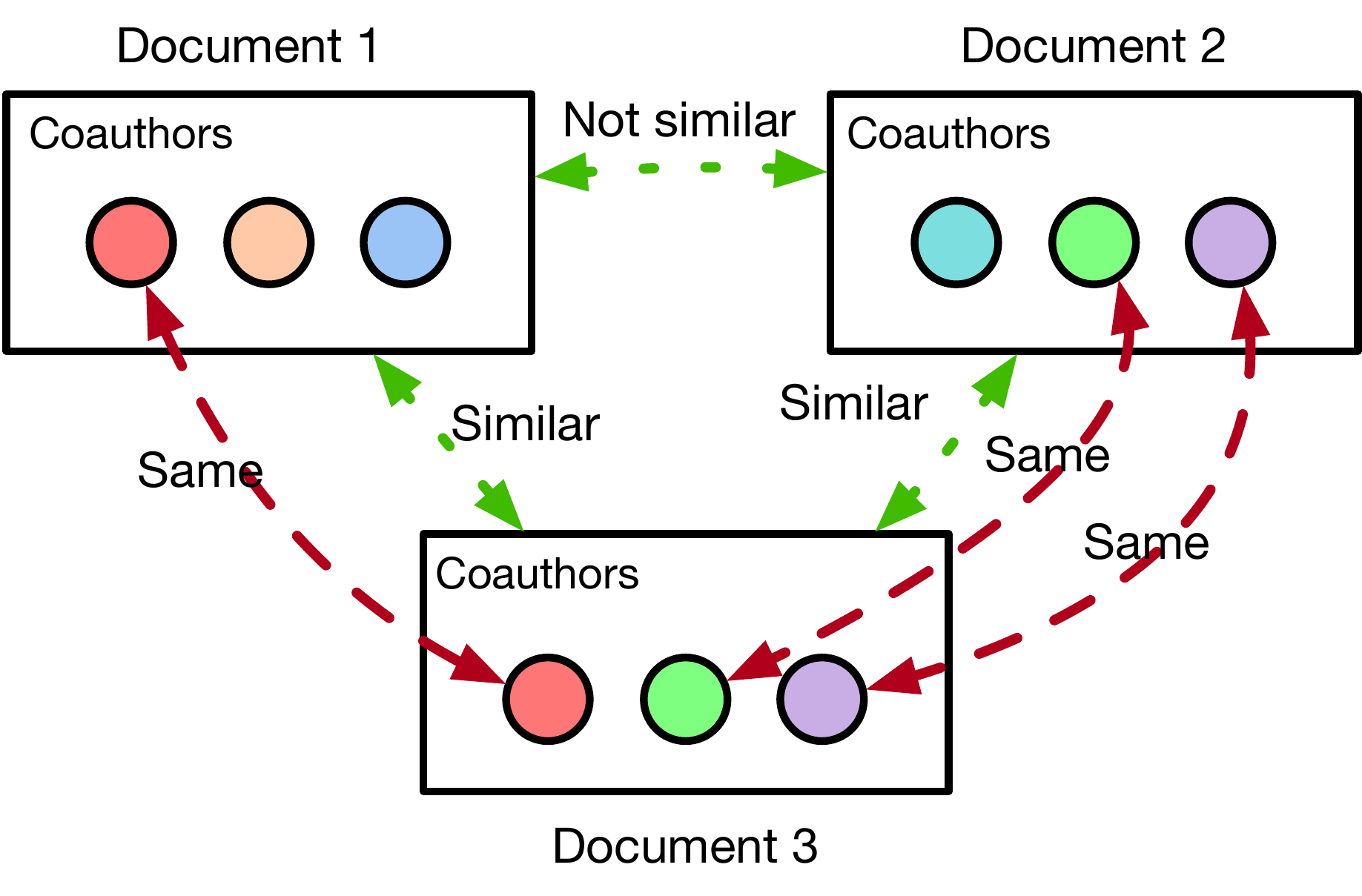}
	\vspace{-0.2in}
	\caption{An ambiguous case in the AND task.}
	\label{fig:ambiguous_example}
\end{figure}

The popularity of the information system has brought explosive growth of academic digital records. The latest estimations show that there are more than 271 million publications, 133 million scholars and 754 million citations on Aminer~\cite{Tang:08KDD} and even larger number of them in Google scholar database. 
\\
Among these digital records, almost all the documents have author name ambiguity problem. Author name ambiguity(AND) means that authors of bibliographic documents are very likely to share identical names with the others so that the scholars' names are not reliable enough to determine their identity. This problem poses a great challenge to the digital bibliographic library, such as Google Scholar, DBLP, Aminer and Microsoft Academic. 
\\
Name disambiguation task is proposed to solve this problem, which is, for a given publication, to distinguish the author who wrote this publication from other authors who share the identical name with him. Despite of great amount of efforts devoted and rapid growth of data-driven methods on graph structure ~\cite{velivckovic2017graph}~\cite{kipf2016semi}~\cite{perozzi2014deepwalk}\cite{kipf2016variational}, the AND problem still remains largely unsolved.
\\
One of the main reason is that existing name disambiguation benchmarks are limited in scale and complexity compared to the great number of academic documents available on the Web. Many machine learning models, such as deep learning models, heavily depend on large-scale and high-quality data. Hence, if we want to propose a sophisticated and robust model to do name disambiguation, a large, diverse and accurate benchmark is indispensable.
\\
However, building such a large-scale benchmark is an extremely challenging task. Annotators need to assign a great number of publications into different clusters. The number of clusters is uncertain and the annotators have to focus on all the pairwise relationships between the publications. For example, if there are 3 thousand (which is a median data size in our dataset) publications to label, the number of pairwise relationships would be 9 million, which is far beyond what humans can handle. Besides, in order to judge whether two publications belong to the same person, it is necessary to consider the relationships between the two publications and other papers. Fig \ref{fig:ambiguous_example} gives an example of this kind of ambiguous cases. Although the document 1 and 2 belong to the same author, their attributes are completely unrelated. It is almost impossible for both human and computers to form a correct judgement on their relationships without the involvement of the document 3. 
Lastly, the workload of labeling each set of data is heavy and it is almost impossible to collaborate, which means, for a given name, each annotator have to complete long-term and challenging process alone and it is hard to merge annotation results from different annotators. This would inevitably lead to poor accuracy and efficiency.
\\
Thus, we propose a supervised inductive name disambiguation method to assist annotators to disambiguate correctly and efficiently. In this method, both supervised and unsupervised methods can serve as similarity models. Based on the constructed similarity graphs, we propose an end-to-end graph neural network model to predict whether two documents belong to the same author and apply community detection algorithm to the generated result. Then, we use visualization techniques to provide display and operation interfaces for annotators. Lastly, our well-designed annotation workflow split annotation process into several parts where annotators' tasks are greatly simplified and annotation results of different annotators could be directly aggregated.
\\
We has utilized this annotation framework to label a great amount of data sampled from AMiner's database, and organized these labeled data as a disambiguation dataset. Our benchmark has significant advantages in scale, complexity and accuracy to the existing ones. Now, the dataset is available online\footnote{\url{https://www.aminer.cn/billboard/whoiswho}}.

\section{Related work}
\subsection{Name Disambiguation methods}
Typically, the author name disambiguation task can be divided into two categories: classification and clustering.
The classification task ~\cite{Han2004Two}~\cite{{Han2005Name}} aims to predict whether two documents refer to the same author or not while the clustering task ~\cite{Tang2012A}~\cite{zhang2017name}\cite{zhang2018name}\cite{Peng2012Disambiguating} is to cluster those documents which belong to the same author together.
\\
Considerable work has been done for these two disambiguation tasks\cite{Wang2011ADANA}\cite{mann2003unsupervised}.
Han~\cite{Han2005Name} defines several similarity functions to evaluate the document similarity based on TF-IDF and NTF, and apply the k-way spectral clustering method on the constructed similarity graphs to disambiguate.
GHOST~\cite{Fan2011On} utilizes the coauthor relationships as input to build a similarity graph and use graph partitioning algorithm to generate the results. 
Tang~\cite{Tang2012A} integrates both document features and graph structural features with the unified probabilistic graphical model HMRF.
Tran~\cite{Tran2014Author} utilize a deep neural network to determine whether two ambiguous documents belong to the same author. 
\\
Recently, utilizing network embedding methods to learn low-dimensional representation for each document is popular, and several state-of-the-art works are based on this approach.
Zhang~\cite{zhang2017name} solve this problem by learning graph embedding from three constructed graphs based on coauthor relationships.
Aminer~\cite{zhang2018name} also leverage a supervised inductive graph neural network model to learn representation for documents and use recurrent neural network to predict the true number of the clusters.
\\
\subsection{Name Disambiguation Datasets}
Either the traditional or the novel, they all need an authoritative data set for evaluation. Supervised methods~\cite{treeratpituk2009disambiguating}~\cite{Tran2014Author}~\cite{zhang2018name} also need labeled data for training and their performance greatly relies on the training data.
\\
Previously, CiteSeerX\footnote{http://clgiles.ist.psu.edu/data/} and Aminer\footnote{https://aminer.org/disambiguation} have published manully-labled author name disambiguation benchmarks respectively. The CiteSeerX dataset consists of 8466 documents with 14 author names while the Aminer dataset consists of 70258 documents with 100 author names. These two benchmarks are widely used to train and evaluate name disambiguation models. However, those models that perform well on these benchmarks not always generalize well in the production environment due to two nontrivial flaws:
\\ 
\vpara{Limited complexity.} The max document numbers of CiteSeerX and AMiner benchmark are 1464 and 999 respectively, which are small compared to the real data scale. In the real world, it is quite common for an author name to refer to thousands documents and some author names, such as Jing Zhang and Wei Wang, even refer to more than ten thousand documents. These huge document sets have much more complicated patterns and higher requirements for the efficiency of disambiguation algorithms. Existing benchmarks can hardly provide evaluations of these aspects.
\\
\vpara{Limited Scale.} Compared to hundreds of millions of papers in the database, the total numbers of documents and author names in these benchmarks are limited. In order to apply the disambiguation models to a production environment, training samples with a larger scale are indispensable.
\\
\vpara{Limited Accuracy.} What we've learned from constructing name disambiguation dataset is that it is a very difficult task for human and they would find great difficulty giving accurate results if there is no well-designed tools. However, for existing AND benchmarks, there is no any specific description of their annotation process, which gives us a reason to worry about the wrong labeling problem.

\subsection{Human-involved Name Disambiguation}
Since there is a gap between the proposed disambiguation models and the real production environment, several human-involved methods are proposed.
\\
D-dupe~\cite{Bilgic2006D} is an interactive framework for entity resolution which visualizes the author collaboration social network into a graph, which helps users to distinguish different persons. 
\\
Shen~\cite{Shen2017NameClarifier} designs several novel visualization interfaces where users can assign newly-coming documents to existing author based on the various information provided by the interfaces.
\\
We take one of Shen's visualization interfaces that cleverly visualizes the coauthor similarity information and document set information as our user interface. We have also made many functional extensions and improvements on it so that it can be used in more complicated annotation task. More details are available in section \ref{sec:appendix}.

\section{properties of \datasetName}
\begin{figure}[t]
     \centering
     \begin{subfigure}[b]{0.4\textwidth}
            \includegraphics[width=\textwidth]{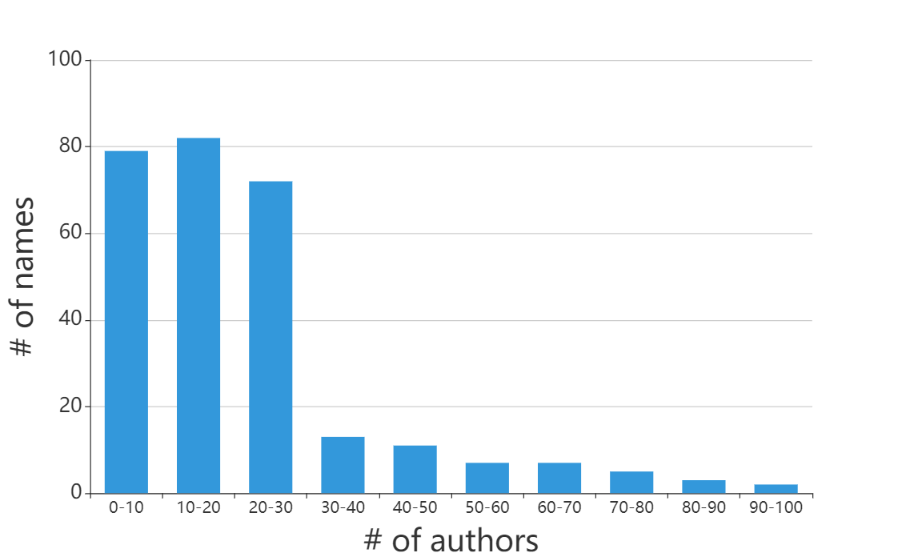}
            \caption{{\small number of authors}}    
        \end{subfigure}
        \hfill
             \begin{subfigure}[b]{0.4\textwidth}
            \includegraphics[width=\textwidth]{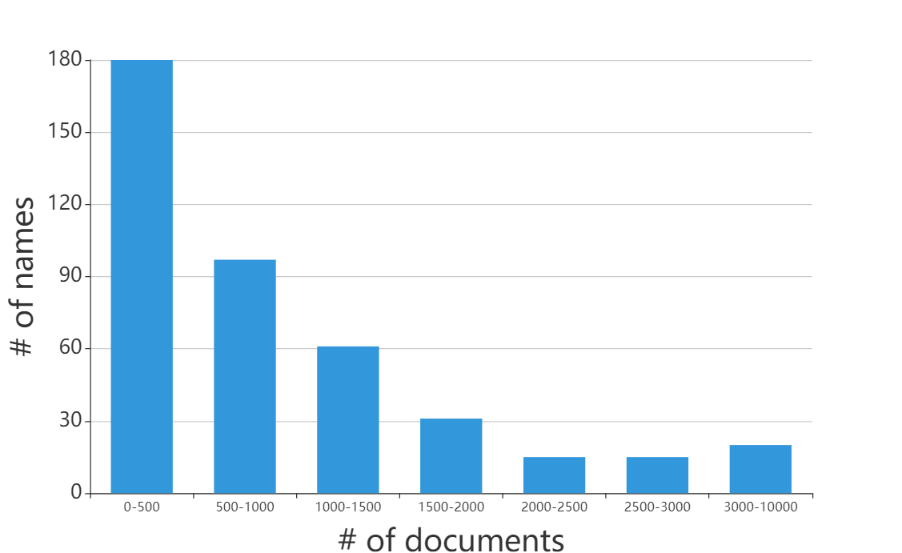}
            \caption{{\small number of documents}}  
        \end{subfigure}
    \caption{Statistics of \datasetName}
    \label{fig:numbers}

\end{figure}
\textbf{Data Annotations.} The raw data of our name disambiguation dataset are collected from the AMiner database as follows: first, we choose the author names based on the number of ambiguous authors and their papers. Then, for each author who shares the ambiguous names, we collect all the papers belonging to the author with their attributes, such as title, abstract, coauthors, affiliations, venues, keywords and so on. Finally, we hire several annotators to label the raw name disambiguation dataset based on our annotation framework.

\textbf{Statistic of \datasetName} To our best knowledge, we have published the world's largest manually-labeled name disambiguation dataset with 399,255 papers belonging to 45,187 persons of 421 common author names Some details of the dataset are listed in the fig \ref{fig:numbers}. 

\textbf{Achievements.} We also organize a data challenge\footnote{https://biendata.com/competition/aminer2019/} based on the published dataset. Moreover, we comprehensively analyze different name disambiguation scenarios and define the two basic task tracks in the challenge: \textit{Name Disambiguation from Scratch} and \textit{Continuous Name Disambiguation}. The challenge was held successfully, which attracted more than 1,000 people formed 500 teams to participate and resulted in some meaningful ideas. The challenge also indicates that the name disambiguation task still remains an open problem and requires further research. To promote the exploration in author name disambiguation field, we plan to further release more labeled data periodically with respective data challenges in the future.

\section{PRELIMINARIES}
In this section, we present the formulation of the name disambiguation annotation task with preliminaries.

\subsection{Task Formulation}
\label{sec:formulation}
\begin{figure}[t]
	\centering
	\hspace*{0cm}
	\vspace*{0cm}
	\includegraphics[width=8cm]{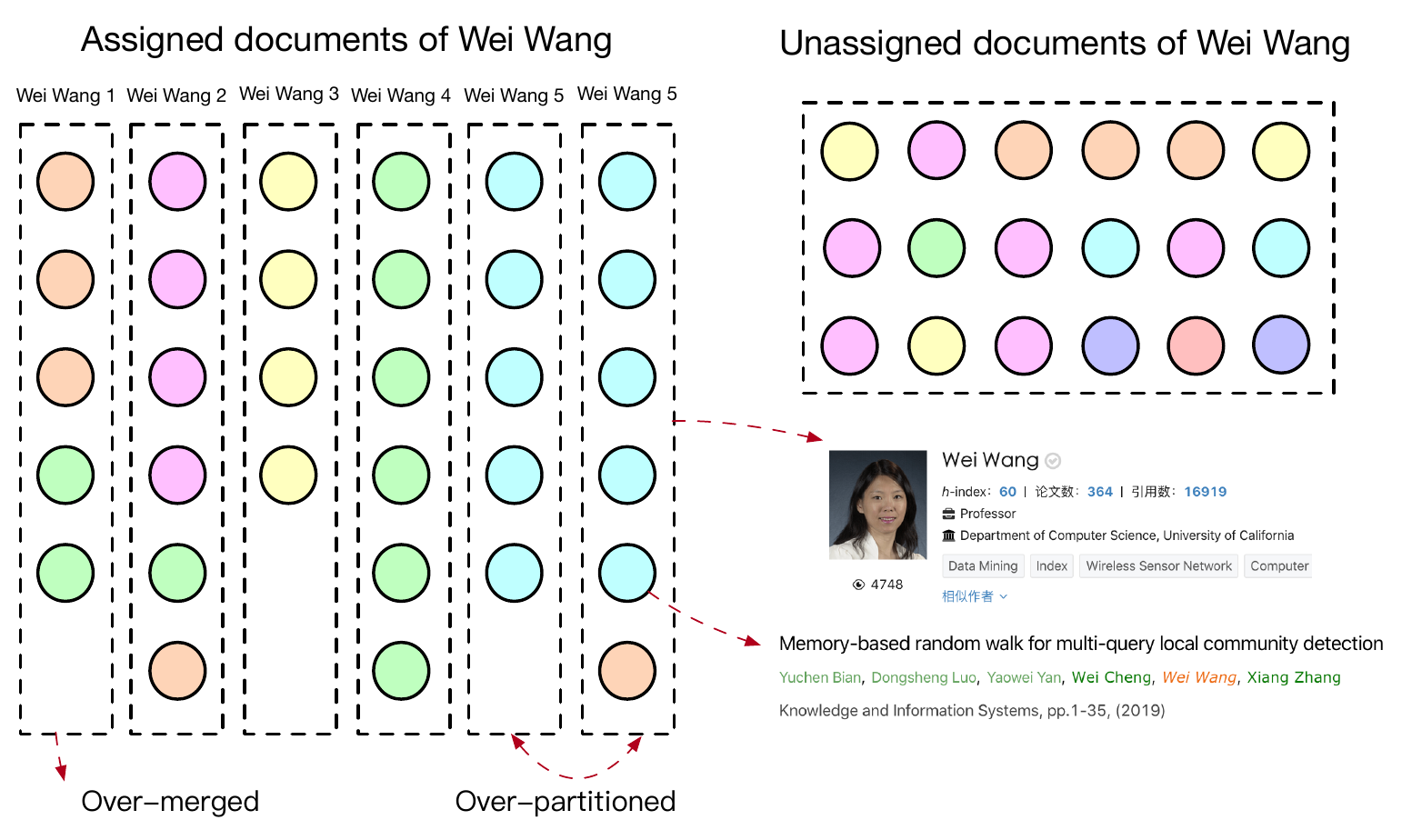}
	\caption{A toy example of input raw data.}
	\label{fig:toy_database}
\end{figure}

Let $a$ be a given name reference, and $\mathcal{D}^a = \{D_1^a, D_2^a, ..., D_N^a \}$ be a set of $N$ documents associated with the author name $a$. 
We call $\mathcal{D}^a$  as the document set of $a$.
We use $\mathbb{I}(D_i^a)$ to denote the identity (corresponding real-world person) 
of $D_i^a$. Thus if $D_i^a$ and $D_j^a$ are authored by the same author, we have $\mathbb{I}(D_i^a) = \mathbb{I}(D_j^a)$.
We omit the superscript in the following description if there is no ambiguity.
Given this, we define the problem of author disambiguation as follows.

\theoremstyle{definition}
\begin{definition}{\bf Name Disambiguation.}
The task of author disambiguation is to find a function $\Theta$ to partition $\mathcal{D}$ into a set of disjoint clusters, i.e., 
$$\Theta(\mathcal{D}) \rightarrow\mathbf{C}, \mbox{where }\mathbf{C} = \{C_1, C_2, ..., C_K\},$$
such that each cluster only contains documents of the same identity---i.e., 
$$\mathbb{I}(D_i) = \mathbb{I}(D_j), \forall (D_i, D_j) \in C_k \times C_k,$$
and different clusters contains documents of different identities---i.e., 
$$\mathbb{I}(D_i) \ne \mathbb{I}(D_j), \forall (D_i, D_j) \in C_{k} \times C_{k'}, k\ne k'.$$
\end{definition}

Our annotation framework begins with existing data in the database, which is more efficient and less challenging by contrast to labelling from scratch.

In addition, documents are integrated into the database in a streaming fashion, hence, there is also a set of unassigned documents $\tilde{C}$ in the input data. Thus we can use $\mathbf{C} = \{C_1, C_2, ..., C_K, \tilde{C}\}$. to denote the origin assignment of the input data.

Figure \ref{fig:toy_database} gives a toy example of raw input data which are sampled from Aminer database.
In Figure \ref{fig:toy_database}, each document has a author named 'Wei Wang'. The color of each document represents its real author identity and the documents within a dashed container denote that they are assigned to the same author profile.
There are typically two types of errors in the assignments:
\begin{itemize}
\item \textbf{Over-merged:} Documents of different authors are wrongly merged into a single profile in the database due to their common attributes or error propagation in the disambiguation process, i.e. 
   $$\exists (D_i, D_j) \in C_k \times C_k, \mathbb{I}(D_i) \neq \mathbb{I}(D_j).$$
  For example, 'W Wang 001' in Figure \ref{fig:toy_database} is over-merged with documents from two different authors. 
\item \textbf{Over-partitioned:} Documents of an individual author are wrongly split as several distinct profiles in the database, i.e.
   $$\exists (D_i, D_j) \in C_k \times C_{k'}, k \neq k', \mathbb{I}(D_i) = \mathbb{I}(D_j).$$
For example, 'W Wang 004' and 'W Wang 005' in Figure \ref{fig:toy_database} are over-partitioned from the same author.
\end{itemize}

For a given author name and its document set $\mathcal{D} = \{D_i\}$, our annotation task has two goals:
\begin{itemize}
    \item 1) generate a perfect document assignment $\mathcal{C} = \{C_i\}$ which not includes unassigned document set $\tilde{C}$.
    \item 2) assign as many documents as possible as long as the first goal is met.
\end{itemize}

\section{framework}
In this section, we present our annotation framework for name disambiguation task. We first introduce the method to extract several similarity graphs from a raw document set. Then we discuss the inductive model which is used to aggregate and refine the similarity graph. With the refined graph, we apply a community detection algorithm to cluster publications which are likely to belong to the same author together. 
\\
After that, the annotators could conduct operations upon these clusters. With well-designed operations and visualization interfaces, annotators are able to complete this difficult task efficiently and correctly. 
\\
Lastly, We discuss the annotation workflow, which decomposes the whole annotation process into four steps: \textbf{cleaning}, \textbf{checking}, \textbf{adding} and \textbf{merging}. During these steps, mutual inspection and majority voting would be applied so that the accuracy of annotation is guaranteed.

\subsection{Building Similarity Graphs}
At first step, we model the pairwise similarity based on several attributes of documents and build multiple similarity graphs for each raw document set. Each similarity graph is referred to a specific document attribute. 

\begin{definition}{\bf Similarity Graph.}
	For a given document set $\mathcal{D} = \{D_i\}$ and a given document attribute $x$, we construct a \textbf{complete weighted} graph as the similarity graph $\mathcal{G}_x(\mathcal{D}) = (\mathcal{D}, \mathcal{E}_x)$. Each edge weight (the edge weight could be zero) is calculated by similarity function $S_x$.
\end{definition}

Both supervised and unsupervised methods can serve as similarity functions. For the name disambiguation model proposed in this paper, we adapt inverse document frequency as the similarity function for its simplicity and scalability. However, in practical deployment, we carefully define the features and utilize Support Vector Machine models to build more accurate similarity graphs. It is mainly because the similarity graphs would be visualized in our annotation interface.


\subsection{Graph Refinement}
In order to cluster documents properly, there should be a single graph where each edge represents the probability that connected documents belong to the same author. Since the similarity of a certain attribute can not serve as this function, it is necessary to aggregate all similarity graphs into a universal graph. We call this process as Graph Refinement.
\\
Actually, we take different graph refinement models as the annotation project progresses. At early stages, we simply sum all the similarity graphs together and use an empirical threshold to filter the edges for simplicity. Because there is limited training data available.
\\


After gaining enough annotation data, we adapt an end-to-end graph neural network model to refine the graph. This model could evaluate the similarity between each pair of documents via topological information from all the similarity graphs so it is possible to solve the tricky problem in fig \ref{fig:ambiguous_example}.
\\
\vpara{Graph Neural Network for Edge Classification}
\begin{figure*}
\centering
  \includegraphics[width=\textwidth]{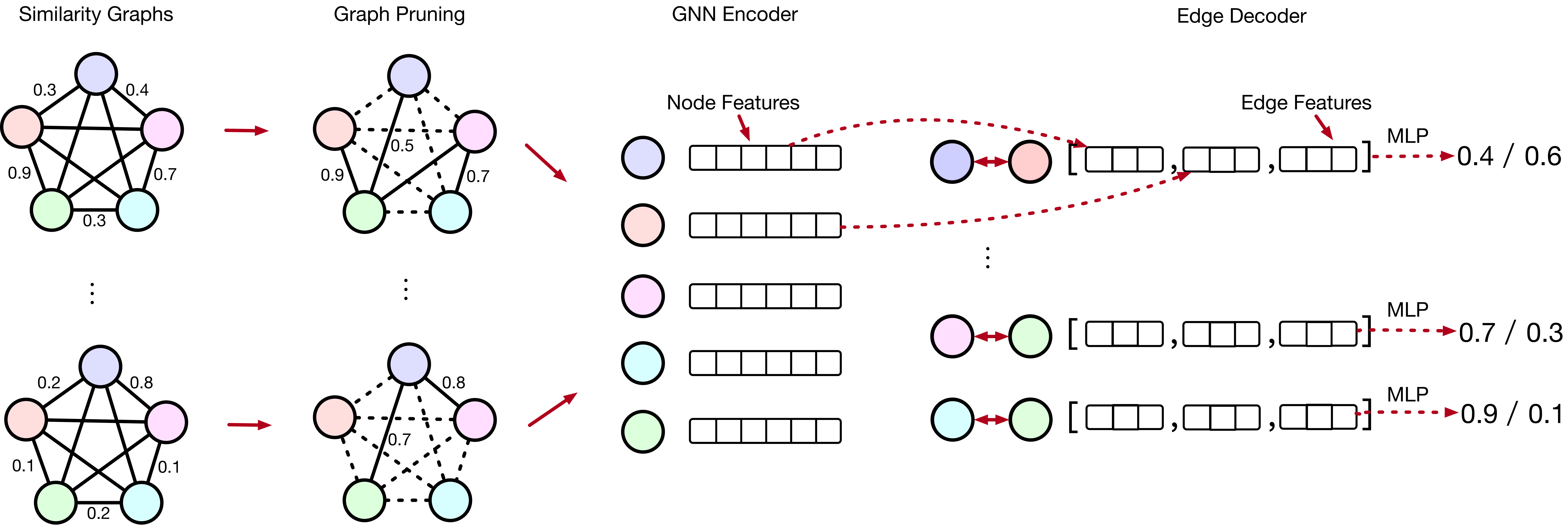}
  \caption{The structure of proposed name disambiguation model.}
  \label{fig:model_structure}
\end{figure*}
Xu~\cite{xu2018powerful} and Hamilton~\cite{hamilton2017inductive} have proven that GNN could learn graph structure information as Weisfeiler-Lehman algorithm if randomize input node features. In that sense, for a given document set $\mathcal{D} = \{d_i\}$ and its similarity graphs $\mathcal{G} = \{\mathcal{G}_x\}$, we generate a random Gaussian vector for each node as input feature. There are two reasons for random initialization:
\begin{itemize}
    \item Almost all valuable information of documents' attribute have been encoded into the similarity graphs.
    \item The involvement of global semantics would significantly increase the complexity, which make model prone to overfitting.
\end{itemize}
Since the similarity graphs are complete graphs and most edges with in them are weak, we apply an adaptive pruning strategy on these graphs.
\begin{definition}
    {\bf Adaptive Graph Pruning.}
    For a given $N \times N$ input adjacent matrix $\hat{A}$, 
    the pruned adjacent matrix $A$ will be produced as:
    $$
    \tilde{A}_{ij} = \begin{cases}
    0 & if \ \hat{A}_{ij} < \frac{\sum{j'}\hat{A}_{ij'}}{N}  \\ 
    \hat{A}_{ij} \ \ & else\\ 
    \end{cases} 
    $$
    $$
    A_{ij} = \frac{\tilde{A}_{ij} + \tilde{A}_{ji}}{2}
    $$
    The pruning strategy would computes a threshold for each row and column\footnote{Each row and column represents the similarity relationships of a specific document}, and the element $\hat{A}_{ij}$ would be filtered by the mean value of row threshold and column threshold. So the symmetry of matrix $A$ is preserved.
\end{definition}
Row normalization is used to normalize the pruned adjacent matrices. $\mathcal{A} = \{A^x\}$ denotes the normalized adjacency matrix set. The adjacency matrices $\mathcal{A}$ are fed into two EGNN~\cite{gong2019exploiting} layers with the node features $V$. The $k^{th}$ layer function and defined as 
$$
V^k = \sigma [\Vert_{x}(A_x V^{k-1} W^{k})]
$$
where $\Vert$ represents the concatenation operation.
Through GNN encoder, the input $N \times N$ adjacency matrices are encoded as N dense vectors $\{V_i\}$, where each vector is referred to a graph node (document). Then we combine all pairs of node feature vectors with corresponding edge feature vector into $N^2$ triplets$(V_i, V_j, E_{ij})$. We concatenate the feature vectors within each triplet and feed them into an MLP classifier to decode the triplets into edges.
For each given document set $\mathcal{D} = \{d_i\}$ and its annotation result $\mathbf{C} = \{C_1, C_2, ..., C_K\}$, we build a complete graph $\mathcal{G}_c(\mathcal{V}, \mathcal{E})$ as the ground truth, where $\mathcal{E} = \{e_{ij}, \forall v_i, v_j \in \mathcal{V} \times \mathcal{V}, i \ne j\} $. According to the results of the annotation, we classify edges into two categories: the positive $\mathcal{E}_p$ and the negative $\mathcal{E}_n$. i.e.
$$
\mathcal{E}_p = \{e_{ij}, d_i, d_j \in C_k \times C_k, C_k \in \mathbf{C}\}
$$
$$
\mathcal{E}_n = \{e_{ij}, d_i, d_j \in C_k \times C_k', k \ne k', C_k, C_k' \in \mathbf{C} \times \mathbf{C}\}
$$
In most instances, the number of negative samples is much larger than the positive, so we apply the weighted loss based on the ratio of positive and negative samples.

\subsection{Sub-clustering}
After graph refinement, we got a refined graph $\mathcal{G}$ for a given document set$\mathcal{D} = \{D_i\}$. The edge $e_{ij}$ between document $D_i$ and document $D_j$ represents how likely is it that $D_i$ and $D_j$ belong to the same author. Based on this graph, we could apply community detection techniques to each assigned document groups $C_i \in \mathcal{C}$ and to unassigned document set $\tilde{C}$ respectively. We call this process as sub-clustering. It splits each assigned group and unassigned set into several sub-groups which the documents within are very likely to belong the same scholar. Hence, the annotators can readily conduct operations upon a batch of documents, which significantly boosts the efficiency of the annotation process.
\\
As long as the annotators conduct an operation, our system would re-cluster the documents. Apparently, efficiency is critical to the online algorithm, so we take \textbf{Speaker-Listener Label Propagation Algorithm}\cite{xie2011slpa}, an improved label propagation algorithm, as our clustering method. The algorithm can form a good result in a short time.
\\
However, the community detection results cannot directly apply to our framework because there will be many huge sub-groups within the results. The huge sub-groups will severely impact visualization effect since we plot the similarity graphs of selected documents. So we set a maximum sub-group size\footnote{The size is empirically set to 50.}, and split all over-size sub-group into several median groups by the \textbf{Breadth-First-Search} method.

\subsection{Annotation Workflow}
\subsubsection{Operation Definition}
According to \S~\ref{sec:formulation}, there are mainly two types of errors: over-merged and over-partitioned. In addition, there is a set of documents $\tilde{C}$ that remains unassigned.
Given these, we define five batch-wise operations as follows:

\begin{itemize}
	\item \textbf{Merge}: A merge operation $\phi_m(C_1, C_2)$ select two assigned groups $C_1$ and $C_2$ and combine all the documents from the two assigned groups into a new assigned group $C_{1,2}$, i.e.
	
    $$\phi_m(C_1, C_2): C_{1,2} \leftarrow \{D_i \in C_1 \cup C_2\}.$$
    
    Intuitively, we prefer to merge assigned groups with similar documents as they are very likely from the same author.
    
	\item \textbf{Separate}: A separate operation $\phi_s(C_k, c_j)$ excludes a sub set of documents $\{D_i \in c_j\}$ from an assigned group $C_k$ and create a new assigned group $C_{j}$ based on $c_j$, i.e.
	
    $$
    \phi_s(C_k, c_j): \begin{cases}
    C_k \leftarrow \{D_i \in C_k \backslash c_j\}, \\ 
    C_j \leftarrow \{D_i \in c_j\}.\\ 
    \end{cases}
    $$
    
    where $c_j \in \Psi(C_k)$. We prefer to perform separate when the documents are similar in $c_j$ but dissimilar to the documents in $C_k \backslash c_j$. 

	\item \textbf{Create}: A create operation $\phi_c(c_j)$ take a sub set $c_j$ from unassigned document set $\tilde{C}$ and create a new assigned group $C_j$, i.e.
	
	$$
	\phi_c(c_j): C_j \leftarrow \{D_i \in c_j, c_j \subset \tilde{C}\}.
	$$
	
	A create operation $\phi_c(c_j)$ will be taken when the documents within $c_j$ are similar (likely from the same author) but dissimilar to any documents in existing assigned groups $\{C_k\}$.
	
	\item \textbf{Assign}: An assign operation $\phi_a(C_k, c_j)$ take a sub set $c_j$ from 
	unassigned document set $\tilde{C}$ and assign these documents into an existing assigned group $C_k$, i.e
	
	$$
	\phi_a(C_k, c_j): C_k \leftarrow \{D_i \in c_j \cup C_k, c_j \subset \tilde{C}\}.
	$$
	
	An assign operation $\phi_c(c_j)$ will be taken when the documents with $c_j$ are similar to the documents in $C_k$ (likely from the same author).
	
	\item \textbf{Exclude}: An exclude operation $\phi_e(C_k, c_j)$  excludes a sub set of documents $\{D_i \in c_j\}$ from an assigned groups $C_k$and set all the documents from $c_j$ as unassigned, i.e.
	
	$$
    \phi_e(C_k, c_j): \begin{cases}
    C_k \leftarrow \{D_i \in C_k \backslash c_j\}, \\ 
    \tilde{C} \leftarrow \{D_i \in c_j \cup \tilde{C}\}.\\ 
    \end{cases}
    $$
	
    where $c_j \in \Psi(C_k)$. We prefer to perform exclude when the documents are dissimilar in $c_j$ and also dissimilar to the documents in $C_k \backslash c_j$. 
	p
\end{itemize}

Any possible disambiguation operation can be converted to a permutation of the above operations.
\subsubsection{Workflow}
We divide annotators' task into four steps: Cleaning, Verifying, Adding and Merging. For each step, we give a clear goal and limited types of operations that annotators can use, which significantly simplify the task so that even the inexperienced can quickly master this annotation task. For each step where annotators can cooperate, the annotators are very likely to give different results. So the different voting strategies would be applied to aggregate the results.
\begin{itemize}
    \item \vpara{Cleaning} This task would be complete independently by a single annotator. The annotator would use \textbf{separate} or \textbf{exclude} operations to clean all the noise in each assigned group. If the annotator cannot make sure if the documents should be removed from the assigned group, the operation should be performed in order to make sure all assigned groups are clean.
    \item \vpara{Verifying} Since the previous step is done by a single annotator, it is necessary to verify the results. So at this step, several annotators get involved. They merely use \textbf{exclude} operation to refine the result given by previous annotator. After verifying, the over-merged problems should be eliminated, which means the over-merged problem does not exist any more and all the documents in the same assigned group belong to the same person.
    \item \vpara{Adding} At this step, several annotators would conduct \textbf{assign} and \textbf{create} operations respectively. In order to maintain high accuracy of labeled data, annotators are asked to perform the operations only if they are quite sure, so the ambiguous documents would remain unassigned. 
    \item \vpara{Merging} At the last step, the annotator would check each pair of assigned groups and decide whether to \textbf{merge} them\footnote{Actually, annotators merely need to check a few pairs, because our visualization interface will filter out those pairs that are totally unrelated}. Putting merge operation as the last step is to avoid the potential problem described in fig \ref{fig:ambiguous_example}. When many documents remain unassigned, it is possible that some over-partitioned document sets seem to be totally unrelated so annotators just skip them. To avoid this situation, it is necessary to conduct \textbf{merge} operation after sufficient documents having been assigned.
\end{itemize}
After applying voting after merging results, we get the annotation result of the final version. Our workflow makes the complicated annotation task simple and efficient while ensuring accuracy.
\subsubsection{Voting Strategy}
There are three steps where the voting strategies are supposed to be applied: \textbf{Verifying}, \textbf{Adding} and \textbf{Merging}. At these steps, each annotator would give a different result. As we mentioned in \ref{sec:formulation}, for a given document set $\mathcal{D} = \{D_i\}$and its assignment $\mathcal{C} = \{C_i\} \cup \tilde{C}$, our ultimate goals is to guarantee the accuracy of $\mathcal{C} \backslash \tilde{C}$. Based on this goal, different strategies are applied at the end of each step.
\begin{itemize}
    \item \vpara{Voting for Verifying} After verifying, each annotator $p^k$ would give a different set of excluded documents $\mathcal{E}^k_i$ for each assigned group $C_i$, where $C_i \subset \mathcal{C} \backslash \tilde{C}$. To maximize the accuracy of $\mathcal{C} \backslash \tilde{C}$, the aggregated result $\mathcal{E}_i$ for assigned group $C_i$ will be the union set of $\{\mathcal{E}^k_i\}$. i.e. $\mathcal{E}_i = \bigcup_k{\mathcal{E}^k_i}$
    \item \vpara{Voting for Assigning (assign operation)} After adding, each annotator $p^k$ would give a set of newly-assigned documents $\mathcal{A}^k_i$ for each assigned group $C_i$. Let $\mathcal{N}(d_j)$ denotes the number of sets $|\mathcal{A}^k_i, d_j \in \mathcal{A}^k_i|$ and $K$ represents the number of annotators, we apply \textbf{majority voting}, so the adding results for each assigned group $\mathcal{A}_i$ are defined as $\mathcal{A}_i = \{d_j, \mathcal{N}(d_j) > K / 2\}$. So the accuracy of assigning is guaranteed. 
    \item \vpara{Voting for Adding (create operation)} On the other hand, after adding, each annotator would also give several sets of documents $\mathcal{C}^k_i$ for \textbf{create} operation. Unlike the \textbf{assign} operation, there would be some conflicts between the results given by different annotators. For better quantification, We formulate the conflicts as the pairwise conflicts. For instance, if an annotator assigns both document $i$ and document $j$ to a newly-created group while another annotator assigns these two documents into two different newly-created groups, there would be a conflicting pair. Otherwise, there would be a verified pair. We apply the voting strategy for \textbf{create} operation base on two pairwise principles:
    1) We merely adopt verified document pairs.
    2) We do not adopt any conflicting document pairs even if they are verified.
    We adapt a greedy search algorithm to merge as many as documents we can without violating any principles above.
    \item \vpara{Voting for Merging} After merging, each annotator $p_k$ would give a set $M_k = \{(C_i, C_j), i \ne j\}$ of assigned group pairs which are going to be merged \footnote{For example, if there are three assigned groups merged together:  $C_i$, $C_j$, $C_k$, there would be three merging pairs: $(C_i, C_j)$, $(C_i, C_k)$ and $(C_j, C_k)$.}.We also apply \textbf{majority voting} to these merging pairs and generate a new set of merging pairs $\mathcal{M}$. Lastly, all the document pairs within the $\mathcal{M}$ would be merged.\footnote{If there are merely two merging pairs (there should be three ideally) between three documents, the three documents would also be merged.} 
\end{itemize}
\subsection{System Efficiency}
For each ambiguous author name, we would model relationships between each pair of documents, Hence, the time complexity of building similarity graphs and graph refinement is \textbf{$O(N^2)$}. Furthermore, the numbers of documents for most ambiguous names are all in thousands, which means that our models need to infer millions or ten millions of times for each ambiguous name. Apparently, it is time-consuming and impossible to deploy online.
\\
To solve this problem, we preprocess the data and cache refined graphs for the names to be disambiguated in advance because refined graph is constant during the annotation process. Only some time-efficient modules, such as community detection and some rendering functions, are deployed online so that the annotator barely feel any delay regardless of the data scale.
\subsection{Visualization Interface}
\section{experiments}
\begin{table*}
\newcolumntype{?}{!{\vrule width 1pt}}
	\newcolumntype{C}{>{\centering\arraybackslash}p{2.2em}}
	\caption{
		\label{tb:performance} Results of Author Name Disambiguation.
		\normalsize
	}
	\centering\small
	\renewcommand\arraystretch{1.0}
	\begin{tabular}{@{~}l@{}|*{1}*{1}{C|CCC|}*{1}{CCC?}{CCC|}*{1}{CCC|}*{1}{CCC}CCC}
		\toprule
		& \multicolumn{1}{c|}{Size}
		&\multicolumn{3}{c|}{Xiao}
		&\multicolumn{3}{c?}{Xiao(F)}
		&\multicolumn{3}{c|}{Louppe~{et al.}}
		&\multicolumn{3}{c|}{Aminer}
		&\multicolumn{3}{c|}{Zhang~{et al.}}
		\\
		\cmidrule{2-4} \cmidrule{5-7} \cmidrule{8-10} \cmidrule{11-13} \cmidrule{14-16} 
		{Name}  & - & {Prec} & {Rec} & {F1} & {Prec} & {Rec} & {F1} & {Prec} & {Rec} & {F1} & {Prec} & {Rec} & {F1} &{Prec} & {Rec} & {F1} \\
		\midrule

Xianghua Li & 303
&95.45	&95.81	&95.63	
&92.40	&90.12	&91.24	
&90.73 	&94.01 	&92.34 	
&95.55 	&98.50 	&97.00  
&99.35	&88.62	&93.68
\\
Xu Shen & 352
&92.38	&77.38	&84.21	
&97.46	&77.32	&86.23	
&88.86 	&16.71 	&28.13 	
&93.54 	&48.71 	&64.06 
&99.03	&90.29	&94.46
\\
Xiaoming Xie & 478
&87.00	&47.01	&61.04	
&71.36	&48.79	&57.96	
&90.86 	&23.28 	&37.06 	
&94.44 	&36.20 	&02.33 
&98.55	&72.38	&83.46
\\
Suqin Liu & 517
&81.71	&86.98	&84.26
&94.04	&88.03	&90.93
&57.82 	&98.04 	&72.74
&83.67 	&70.80 	&76.70 
&98.77	&48.36	&64.93
\\
Makoto Inoue & 661
&93.32	&79.75	&86.00	
&80.32	&74.69	&77.40	
&95.03 	&52.09 	&67.29 	
&98.55 	&81.02 	&88.93 
&98.75	&92.14	&95.33
\\
Jihua Wang & 713
&93.56  &86.20	&89.24	
&86.40	&75.48	&80.57
&78.96 	&84.95 	&81.84 	
&94.90 	&45.70 	&61.69 
&98.46	&44.73	&61.51
\\
Chao Deng & 745
&95.04	&87.10	&90.89	
&85.44	&78.57	&81.86	
&97.01 	&77.80 	&86.35 	
&99.08 	&72.14 	&83.50 
&98.99	&86.53	&92.34
\\
Qiang Wei & 1130
&92.78	&67.67	&78.26	
&83.02	&74.62	&78.60	
&97.93 	&51.62 	&67.61 	
&97.49 	&34.08 	&50.51 
&99.11	&55.43	&71.10
\\
Xiaohua Liu & 1334
&97.11	&96.81	&96.96	
&94.80	&96.49	&95.63	
&95.38 	&96.06 	&95.72 	
&99.41 	&73.08 	&84.23 
&97.26	&73.47	&83.71
\\
Weimin Liu & 1484
&83.17	&77.89	&80.44	
&76.73	&81.74	&79.15	
&92.94 	&33.80 	&49.57 	
&97.93 	&21.45 	&35.19 
&97.84	&50.31	&66.45
\\
Min Yang & 2244
&92.11	&61.23	&73.56	
&75.86	&61.67	&68.03	
&96.85 	&34.34 	&50.70 	
&98.54 	&31.14 	&47.32 
&98.18	&38.59	&55.41
\\
Jing Li & 4950
&94.62	&88.83	&91.64	
&72.64	&81.18	&76.67	
&99.44 	&56.00 	&71.65 	
&99.61 	&51.41 	&67.82 
&98.56	&75.61	&85.57
\\
Jing Zhang & 6141
&94.91	&73.49	&82.84	
&65.41	&50.59	&57.06	
&98.41 	&33.61 	&50.11 	
&98.99 	&25.97 	&41.14 
&95.78	&25.29	&40.02
\\
\hline
Micro Avg. & -
&93.69	&75.65	&\textbf{83.71}	
&72.3	&68.16	&70.17	
&97.81 	&50.40 	&66.53 	
&98.96 	&44.50 	&61.39 
&98.24	&63.04	&76.80
\\
Macro Avg. & - 
&93.27 	&71.67 	&\textbf{80.39} 	
&78.83 	&64.34 	&69.68
&90.58 	&51.83 	&60.72 	
&96.38 	&57.80 	&69.97 
&97.87	&67.20	&78.11 
\end{tabular}
	
\end{table*}

We conduct several comprehensive experiments to evaluate the accuracy of our proposed annotation framework. We also evaluate recent state-of-the-art methods ~\cite{zhang2018name}~\cite{zhang2017name}~\cite{louppe2016ethnicity} for author name disambiguation task on our benchmark. Lastly, we compare these state-of-the-art methods with ours to demonstrate its superiority over those methods.
\subsection{Name Disambiguation Evaluation}
\subsubsection{Experiment Datasets}
We sampled 320 author names from our dataset, and split them into 200, 60, 60 for training, validation and testing. Each author name refers to a totally different document set. 
\subsubsection{Experiment Setting}
We evaluate three state-of-art name disambiguation methods on our dataset.
\\
\vpara{Aminer~{et al.}~\cite{zhang2018name}:} This method learns a supervised inductive embedding model based on manually-labeled data, and use an unsupervised graph auto-encoder to refine the embedding on the local linkage graph constructed based on the common features between documents. 
\\
\vpara{Zhang~{et al.}~\cite{zhang2017name}:} The second method is an unsupervised method which constructs three graphs (document-document, author-author and document-author) based on coauthors use the triplets sampled from graphs to optimize graph embedding.
\\
\vpara{Louppe~{et al.}~\cite{louppe2016ethnicity}:} This semi-supervised method first trains a pairwise distance function base on a set of carefully designed similarity features. Then a semi-supervised HAC algorithm is used to determine clusters.
\\
Our method is indicated by \textbf{Xiao}. In our method, we leverage both pairwise document similarity information and topological information of similarity graphs to predict whether two document belong to the same person. So we also present the performance of topological component to further analyze the contribution of edge features.
\\
\vpara{Xiao(F)}
This method merely use the topological information of similarity graphs, where the decoder merely takes document embedding learned from GNN model as input.
We evaluate these disambiguation models by the pairwise Precision, Recall and F1-score on 13 sampled author names. We also use both micro and macro averaged scores to evaluate overall performance of each method. The averaged scores are calculated on the complete testing set.
\subsubsection{Experiment Results}
Table \ref{tb:performance} shows the performance of different disambiguation methods on sampled test names of different sizes. According to the results, the performance of both Aminer and Zhang are not ideal on the large document sets since their micro average score is much lower than the macro one. Benefiting from a scalable end-to-end training method, our method (Xiao) outperform the other state-of-the-arts in both macro and micro average score (+17.18\% and +19.67\% over Louppe, +22.32\% and +10.42\% over Aminer, +6.91\% and +2.28\% over Zhang). And the results also show the feeding edge features directly into decoder could greatly boost our performance (+13.54\% and +10.71\%). However, there is still a huge gap between our model and human.
\subsection{Framework Analysis}
\begin{figure}[t]
	\centering
	\includegraphics[width=9.0cm]{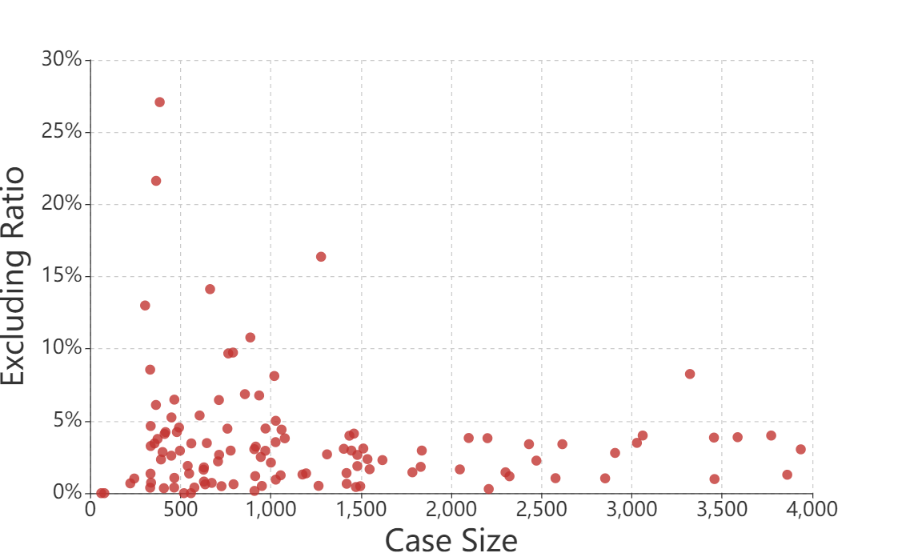}
	\vspace{-0.2in}
	\caption{Excluding Ratio of Verifying}
	\label{fig:excluding_ratio}
\end{figure}
We further investigate the performance of annotators during the annotation process. During the \textbf{Verifying} step, a noisy disambiguation result made by a single annotator would be refined by multiple annotators with our voting strategy so all seemingly wrongly-assigned documents would be excluded. We count the numbers of these excluded documents $E$ and the numbers of the rest assigned documents $R$ during the annotation process. The numbers are shown in fig \ref{fig:excluding_ratio}. The excluding ratio $ER$ is computed as $ER = \frac{E}{R + E}$, which is the proportion of seemingly wrongly-assigned documents and indicates the accuracy of result made by a single human annotator in our annotation framework. In fig \ref{fig:excluding_ratio}, it is shown that the excluding ratios of most document sets are less than $5\%$, which indicates, with our annotation framework, a human annotator can complete a disambiguation task with less than $5\%$ wrongly-assigned documents in most instances. Besides that, we also found, for sampled document sets whose sizes range from 0 to 4000, the distribution of excluding ratios is stable, which further demonstrates that our annotation framework is scalable so that we can apply our annotation framework to label larger and more complex document sets.
\begin{figure}[t]
	\centering
	\includegraphics[width=9.0cm]{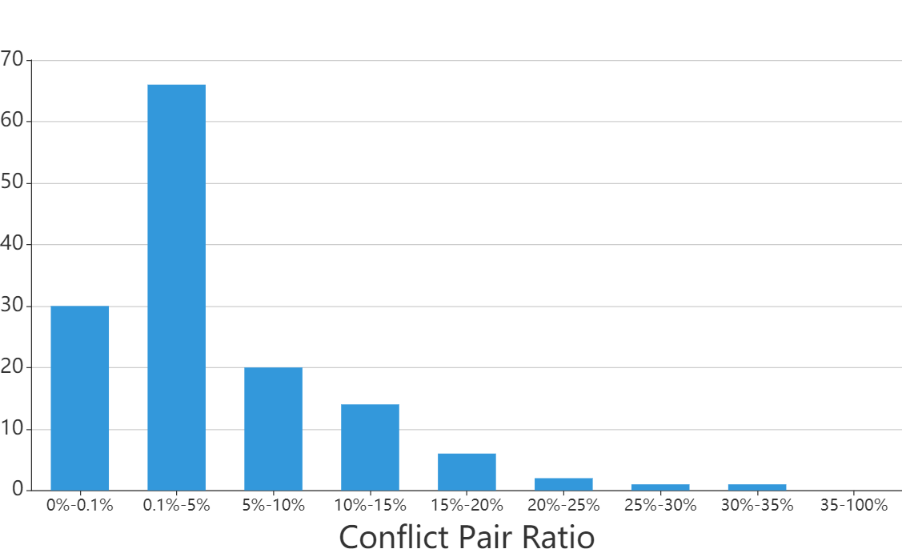}
	\vspace{-0.2in}
	\caption{Conflict Pair Ratio of Create}
	\label{fig:conflict_ratio}
\end{figure}

As we mentioned before, there would be pairwise conflicts between the results given by different annotators during the \textbf{adding} step. We define conflicting pair ratio $CPR$ in the following way. For two annotators $i$ and $j$, let $N_{ij}$ be the number of common documents they both conduct \textbf{create} operation to and $C_{ij}$ represents the number of conflict pairs between them, conflict pair ratio is calculated as $CPR = \overline{CPR_{ij}}$, where $CPR_{ij} = \frac{C_{ij}}{{N_ij}^2}$. Fig \ref{fig:conflict_ratio} shows the distribution of $CPR$ we have counted during the annotation. According to fig \ref{fig:conflict_ratio}, the creating conflicts widely exists while their scales are generally small, which indicates that annotators are very likely to make minor mistakes and seldom make big ones in our framework. This kind of mistakes can be detected and completely eliminated by our workflow, which further demonstrates the accuracy of our dataset.

In the real annotation scenario, we arrange three annotators for each collaboration step, which means the amount of work triples in these collaboration steps. Our records show that it would take about 300 man-hours to label 100,000 documents under this arrangement. On average, each person can label at least 600 to 700 documents per hour with the assistance from our annotation system.

\section{conclusion}
The experiment codes and annotation system demo are available online.\footnote{\url{https://www.aminer.cn/annotation}}
In this paper, we dived into the issues of Author Name Disambiguation.
\\
First, we propose a novel crowdsourcing framework to build world's largest manually-labeled Author Name Disambiguation dataset, \textbf{WhoisWho}, which provides a new point of view for researchers. 
Through comprehensive evaluation and analysis, we demonstrate that our annotation work is highly efficient and annotation result is accurate. We also organized a competition based on the published dataset and resulted in some meaningful ideas.
\\
On the other hand, we adapt an inductive supervised model for AND task and apply it into our annotation framework to assist annotators. We evaluate it with several state-of-the-arts name disambiguation methods in our benchmark. Experiment results demonstrate the advantage of our method over stat-of-the-art author name disambiguation methods. The results also show that even the best name disambiguation model still has a huge gap with human, which indicate AND is still an open problem and there is still large space for AND methods to improve.
\\
In the future, we will continue to expand the scale of annotation data and release the annotation task to the Internet in the form of crowdsourcing, so that more people can participate in our annotation task. New annotation results will be available with new data challenges soon.
\\
In addition to the manually-labeled data, several works\cite{abdulhayoglu2017use}\cite{schulz2014exploiting} evaluate their models with the certified information in Google Scholar. This kind of data is very tiny for one author name, so it would not be used as a dataset in most cases. However, this data has great advantages in accuracy and diversity. In the future, We will try to use this data to verify accuracy of our annotation results.
\section{Acknowledgement}

\bibliographystyle{ACM-Reference-Format.bst}
\balance
\bibliography{reference.bib} 

\newpage
\newpage
\section{Appendix}
\label{sec:appendix}
\begin{figure*}
\centering
  \includegraphics[width=\textwidth]{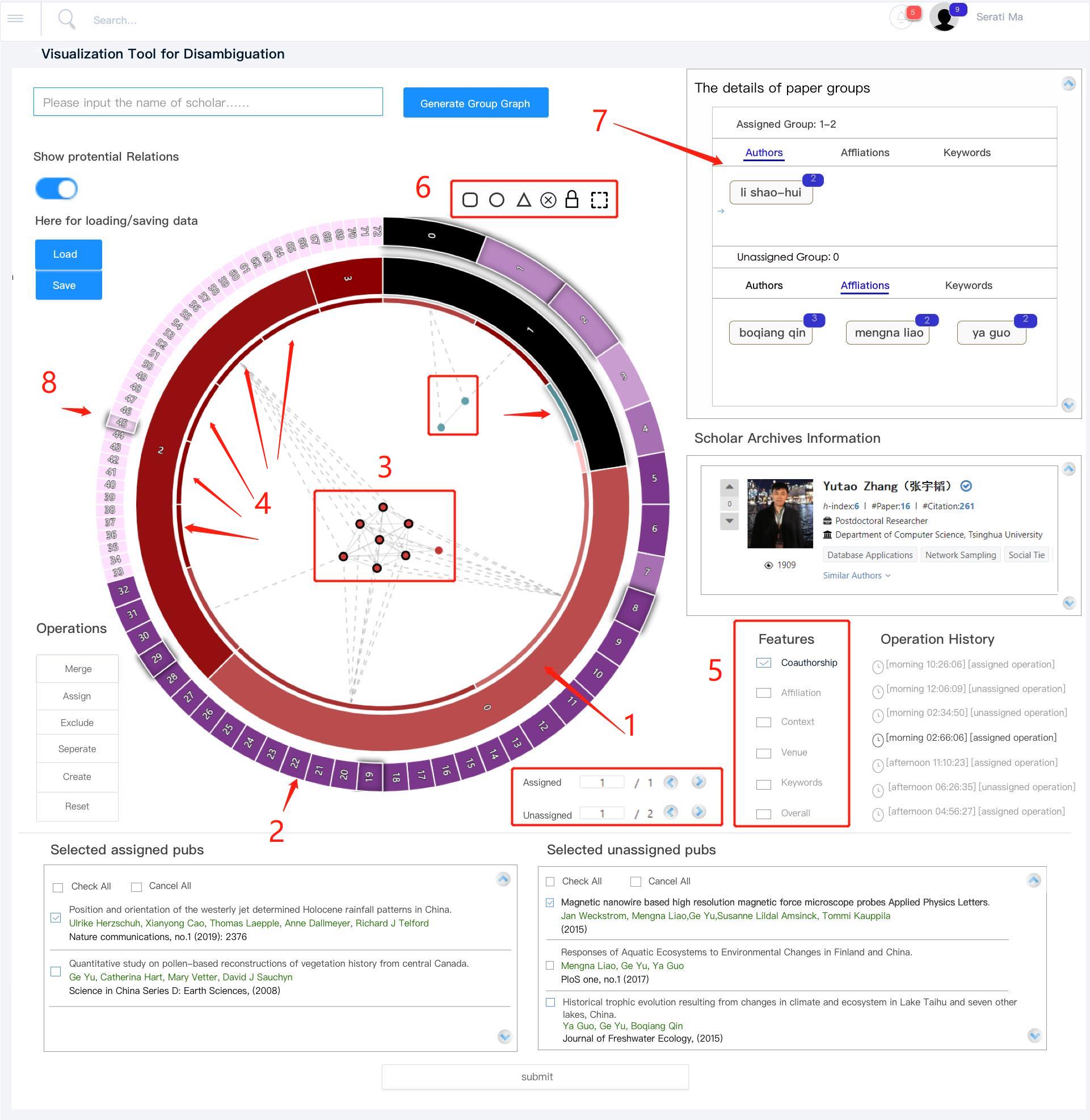}
  \caption{The user interface of proposed annotation system}
  \label{fig:interface}
\end{figure*}

\subsection{User Interface}
In this section, we will introduce our user interface, which is shown in the Fig \ref{fig:interface}. The prototype of this visualization interface is proposed by Shen. We will first introduce the characters designed by the previous work, and then introduce the improvements we have made on it. We will follow the order marked in the figure \ref{fig:interface} one by one.

\subsection{Assigned Groups}
The hollow circle pointed by the arrow 1 represents the assigned document groups and each segment of it represents a certain ambiguous group. If a certain segment is clicked, the profile information of this group would be displayed on the left side of the interface and the documents belonging to this group would be visualized at the center of the circles. Each group size is encoded as the segment length while the group quality is encoded as the segment's color. The darker the segment color is, the more similar the papers in the group are.
\subsection{Unassigned Groups}
The other hollow circle pointed by the arrow 2 represents the unassigned document groups. Each segment of it represents a sub-clustering group. The characters of this component is exactly the same as the previous one.
\subsection{Documents and Relationships}
Each document belonging to the selected groups would be plot as a graph node in the center of circles. If two documents have any common authors, there would be an edge between the corresponding nodes. Besides that, there are some edges existing between the documents and the assigned groups, which means that there is at least one document within the assigned group happening to have common authors with the connected documents. The links between the groups and documents are called potential links.
These visualization technologies above are proposed by the previous work. Next, we will introduce our original new features.
\subsection{Sub-groups of Assigned Groups}
Our framework would use sub-clustering method to split each assigned document group into several sub-groups. The segments pointed by arrow 4 indicate the sub-groups of assigned group 3. These segments share the same features with previous two and facilitate annotators to conduct more precise operations.
\subsection{Features Selection}
In the previous work, the edges between the documents and groups indicate author similarity. Since our annotation framework would take various document attributes into consideration, we extend the functions of edges to enable them to represent more complex relationships. The annotators can choose multiple document features which they are interested in. If two documents are connected, it means that at least one of selected attributes are similar.
\subsection{Node Interfaces}
Our system provides several node interfaces to meet the requirements from annotators. The marking interface allow annotators to mark nodes by modifying their shapes. The freezing interface could fix the node positions while brush interface enable them to select a group of nodes quickly.
\subsection{Groups Info Extraction}
For each sub-group, our framework would count the frequency of features and plot them separately. The annotator can analyze each selected group quickly or simply click a feature to select all the corresponding documents.
\subsection{Ambiguous Potential Links}
The previous work designs a kind of potential links between the assigned groups and documents, which help annotator quickly find the assigned groups they are interested in. We extend this design to unassigned document sets, we use the way of adding shadows to point it out so the annotator can quickly find target unassigned groups or documents as well.

\end{document}